\documentclass[12pt]{amsart}
\usepackage{comment,amssymb,latexsym,cite,upref}
\usepackage{finalT}

\numberwithin{equation}{section}

\newcommand{\id}{\mathord{{\mathrm 1}\kern-0.27em{\mathrm I}}\kern0.35em}
\newcommand{\del}{\partial}

\newcommand{\AND}{{\quad\text{and}\quad}} 

\newcommand{\fst}{1^{\text{st}}}
\newcommand{\snd}{2^{\text{nd}}}
\newcommand{\aph}{\ensuremath{\textstyle{\frac{\alpha'}{2}}}}
\newcommand{\aphsq}{\ensuremath{\textstyle{\frac{{\alpha'}^2}{2}}}}
\newcommand{\apq}{\ensuremath{\textstyle{\frac{\alpha'}{4}}}}

\newcommand{\itap}{\ensuremath{\textstyle{\frac{2}{\alpha'}}}}
\newcommand{\ifap}{\ensuremath{\textstyle{\frac{4}{\alpha'}}}}

\newcommand{\Half}{\ensuremath{\textstyle\frac{1}{2}}}

\newcommand{\norm}[1]{\|#1\|}

\newcommand{\Mcl}{\mathcal{M}}

\newcommand{\Rc}{\mathcal{R}}

\newcommand{\gt}{\tilde{g}{}}

\newcommand{\Rtl}{\tilde{R}{}}

\newcommand{\Deltat}{\tilde{\Delta}{}}

\begin{document}

\title[The $\snd$ order RG flow in 2D]{The $\snd$ order renormalization group flow for non-linear sigma models in 2 dimensions}
\author[T.A. Oliynyk]{Todd A. Oliynyk}
\address{School of Mathematical Sciences\\
Monash University, VIC 3800\\
Australia}
\email{todd.oliynyk@sci.monash.edu.au}
\begin{abstract}
We show that for two dimensional manifolds $M$
with negative Euler characteristic there exists
subsets of the space of smooth Riemannian metrics
which are invariant and either parabolic or
backwards-parabolic for the $\snd$ order RG flow.
We also show that solutions exists globally on
these sets. Finally, we establish the existence
of an eternal solution that has both a UV and
IR limit, and passes through regions where the
flow is parabolic and backwards-parabolic.
\end{abstract}

\maketitle

\sect{intro}{Introduction}

The world-sheet nonlinear sigma model renormalization group
flow arises from quantizing the classical action
\eqn{action}{
S(x)=\frac{1}{4\pi \alpha ^{\prime }}\int_{\Sigma }\gamma ^{\alpha \beta
}g_{ij}(x)\partial _{\alpha }x^{i}\partial _{\beta }x^{j}dV(\gamma ),
}
where $\alpha ^{\prime }>0$ is the string coupling constant, $(\Sigma
,\gamma )$ is a $2$-dimensional Riemannian manifold (i.e. world sheet), $%
(M,g)$ is a $n$-dimensional Riemannian manifold (i.e. target space), and $x$
$:\Sigma $ $\rightarrow $ $M;$ $(\theta ^{1},\theta ^{2})$ $\mapsto $ $%
(x^{1}(\theta )$ $,\ldots $ $,x^{n}(\theta ))$ is a map. A perturbative
quantization of this classical theory requires the introduction
of a  momentum cutoff $\Lambda >0$, and gives rise to a one
parameter family of quantum field theories indexed by $\Lambda$.
The requirement that the family of quantum field theories be equivalent on
length scales $L\gg 1/\Lambda $ leads to
the renormalization group (RG) flow equations
\eqn{fullrg}{
\del_\Lambda g_{ij}=-\beta _{ij}^{g}\,.
}
In the regime where perturbation theory is valid $(\alpha'\ll 1)$, the
$\beta $
-functions $\beta _{ij}^{g}$ can be expanded in powers of $\alpha ^{\prime }$
\cite{Frie85,JJM89}:
\eqn{beta}{
\beta _{ij}^{g}=\alpha'R_{ij}+\aphsq
R_{iklm}R_{j}{}^{klm}+\mathrm{O}({\alpha ^{\prime }}^{3})\,.
}
Defining a ``time'' by $t=-\ln (\Lambda )$, the RG flow equations
become
\leqn{rgexp}{\del_t g_{ij}=-\alpha'R_{ij}-\aphsq R_{iklm}R_{j}{}^{klm}+
\mathrm{O}({\alpha'}^{3}).}
It is expected that in the perturbative regime, the $\fst$ order
truncation
\leqn{forg}{
\del_t g_{ij}=-\alpha'R_{ij}
}
should provide a ``good approximation'' to the full flow. However,
without an estimate of the error, the notion of a good approximation
cannot be quantified. The problem of understanding the error is
obstructed by the fact that a mathematically rigorous
quantization of the non-linear sigma model is presently unavailable.
However, if it ultimately turns out that expansion \eqref{rgexp} obtained
using perturbation theory is valid for the RG flow,
even as an asymptotic expansion,
then it is not unreasonable to expect that the error between the full flow
and \eqref{forg} will be qualitatively described by the $\snd$
order truncation
\leqn{sorg}{
\del_t g_{ij}=-\alpha'R_{ij}-\aphsq R_{iklm}R_{j}{}^{klm},
}
at least for situations where the curvature is not too large. We note
that this expectation is borne out in other field theories
where it has been established that it is enough to consider the
$\snd$ order truncation of the RG
flow to establish the existence of a continuum limit \cite{GK85}. This
reinforces the view that knowing the $\fst$ order flow is not
always enough for applications where quantitative control on the error
is required.

As has been noted now many times, the $\fst$ order RG flow \eqref{forg}
coincides with Ricci flow.
It is known that there are many solutions to Ricci flow that become
singular at a finite time. As shown by Hamilton \cite{Ham82}, a singular
time $T$ of Ricci flow is characterized
by curvature blow up: $\lim_{t \nearrow T} \|R_{ijkl} R^{ijkl}\|_{L^\infty(M)} = \infty$.
This suggests that near a singular time for Ricci flow, the higher
order curvature corrections in \eqref{rgexp} will dominate the behaviour
of the flow even for $\alpha' \ll 1$. Thus, it is natural from
this viewpoint to  consider the higher order truncations
of \eqref{rgexp} to try and capture the effect of higher order curvature
terms.

With the above motivation in mind, the main aim of this article is
to continue the study initiated in \cite{CGTO} of the $\snd$ order
RG flow \eqref{sorg}
using techniques from geometric
analysis. To facilitate comparisons with Ricci flow,
we rescale the time and metric to bring \eqref{sorg} into the form
\leqn{rg}{
\del_t g_{ij} = -2 R_{ij}-\aph R_{iklm}R_j{}^{klm}
}
which makes the leading term  consistent with the standard presentation
of Ricci flow.

Although Ricci flow may be recovered in the limit $\alpha'\searrow 0$, the
$\snd$ order RG flow differs from Ricci flow in two important respects:
it is fully non-linear, and it is not parabolic for all choices of
$\alpha'$ and
$g_{ij}$. Therefore, in addition to curvature blow up, loss of
parabolicity along the flow presents a possible new mechanism
for singularity formulation.
To investigate this possibility, we restrict ourselves to the simplest possible
setting of a closed two dimensional target space $M$.
In this case, the curvature tensor
is given by
\eqn{creature}{
R_{ijkl} = \Half R\bigl(g_{il}g_{jk} - g_{ik}g_{jl}\bigr)
}
which implies that the $\snd$ order RG equations
\eqref{rg} reduce to
\leqn{rg2d}{
\del_t g_{ij} = - \Rc g_{ij},
}
where
\leqn{Rc}{
\Rc = R + \apq R^2 .
}

Following the Ricci flow terminology, we will call a solution $g(t)$ to \eqref{rg}
(equivalently \eqref{rg2d}) \emph{ancient}, \emph{immortal}, or \emph{eternal},
if the solution exists on a time interval of
the form $-\infty < t < t_0$, $t_0 < t < \infty$, or $-\infty < t < \infty$,
respectively. We will also use the physics terminology of a \emph{UV}
(\emph{IR}) limit which refers to a fixed point of \eqref{rg} from which
an ancient (immortal) solution originates (terminates).
From both a mathematical and physical viewpoint, the existence and classification of
the ancient, immortal, and eternal solutions are of fundamental interest. For
physical applications, the UV limits are of particular importance
as they correspond to cut-off removal for the quantum field theory. In other words,
a UV limit identifies quantum field theories with a well defined microscopic limit.
The IR limits also have a physical interpretation and correspond to
a well defined macroscopic limit.

The main result of this paper is to show that if $M$ has negative Euler
characteristic, then there exist large regions
in the space of smooth Riemannian metrics $\Mcl$ on $M$
for which the flow is either parabolic or backwards-parabolic, and that these
regions remain invariant under the flow. Moreover, on these invariant
regions, the flow has good long term existence properties.

We also establish the  existence of an eternal solution to \eqref{rg} with
both a UV and IR limit that passes from a region
of backward-parabolicity into a region of parabolicity.
We find this solution particularly interesting as it is a consequence
of the curvature correction $\aph R_{iklm}R_j{}^{klm}$ term to Ricci flow.
This solution shows that the lack of uniform parabolicity for all
choices of $\alpha'$ and $g_{ij}$
should not necessarily be viewed as a defect.
Instead, the notion of uniform parabolicity should be replaced with
that of invariant parabolic
or backwards-parabolic sets. As this
paper demonstrates, the existence of neighbouring parabolic and
backward-parabolic regions opens up the possibility for constructing
solutions by joining together two solutions at a degenerate
parabolic point that joins a backward-parabolic region to
a parabolic one.

\sect{rgeqns}{Parabolicity of the $\snd$ order RG equations}

Due to the conformal nature of the $\snd$ order RG flow \eqref{rg2d}, it is consistent given initial data $g_{ij}\bigl|_{t=0} = \gt_{ij}$
to write $g_{ij}(t) = e^{u(t)}\gt_{ij}$ in which case the initial value problem 
\leqn{rg2divp}{
\del_t g_{ij} = -\Rc g_{ij} \quad : \quad g_{ij}(0) = \gt_{ij}
}
is equivalent to 
\leqn{rg2u}{
\del_t u = -\Rc(u) \quad : \quad u(0)= 0,
}
where
\leqn{cR}{
\Rc(u) = R+\apq R^2 \AND R = e^{-u}\bigl(-\Deltat u + \Rtl\bigr). 
}
Here, $\Rtl$ and $\Deltat$ denote the Ricci scalar and Laplacian of
$\gt_{ij}$, respectively. 

The linearization 
\eqn{par1}{
D\Rc(u)\cdot v = \bigl(1+\aph R\bigr)\bigl(-e^{-u}\Deltat v -
R v \bigr),
}
shows that \eqref{rg2u} is \emph{parabolic} when
$1+\aph R > 0$ and \emph{backwards parabolic} when
$1+\aph R < 0$.
This and the properties of the evolution equation to be
described in the next section motivate us to define the following subsets of
the space of smooth Riemannian metrics $\Mcl$:
\leqn{Mpm}{
\Mcl_+ = \{\, g\in \Mcl \, | \, - \itap < R < 0\, \}
\AND \Mcl_-  = \{\, g\in \Mcl \, | \, -\ifap < R < - \itap\, \}
}
To avoid the situation where these sets are empty, we will,
as mentioned in the introduction, restrict ourselves to
closed manifolds with negative Euler characteristic.

\sect{gexist}{Invariance of $\Mcl_\pm$ and global existence}

\begin{thm} \label{thmp} \mnote{[thmp]}
Suppose $\gt \in \Mcl_+$. Then there exists a smooth one
parameter family of metrics
$g(t)$ for  $0\leq t < \infty$ that satisfy the following:
\begin{itemize}
\item[(i)] $\; g(t)$ solves the $\snd$ order RG equation \eqref{rg}
with $g(0)=\gt$,
\item[(ii)]$\; g(t) \in \Mcl_+$ for all $t\geq 0$, and
\item[(iii)] there exists a constant $C_{\Rtl}$ depending
only on $C^+_{\Rtl} =  \max_{x\in M}\Rtl(x)$ and
$C^-_{\Rtl} = \min_{x\in M}\Rtl(x)$ such that
\eqn{thmp1}{
C^-_{\Rtl} \leq R(t,x) \AND  |R(t,x)| \leq \frac{C_{\Rtl}}{1+t}
}
for all $(t,x) \in  [0,\infty)\times M$.
\end{itemize}
\end{thm}
\begin{proof}
Since the equation \eqref{rg2u} is parabolic whenever
$g = e^u\gt \in \Mcl_+$, it follows by standard local existence
theorems for parabolic equations (see Proposition 8.1, pg. 338 of \cite{TayIII} ) that there exists a
smooth solution $u(t)$ to the initial value problem \eqref{rg2u} defined
on some interval $0\leq t < T$. Given this solution, a short calculation
using \eqref{rg2d} shows that the Ricci scalar satisfies the
equation
\leqn{thmp2}{
\del_t R = \Delta \Rc + \Rc R ,
}
or equivalently
\leqn{thmp3}{
\del_t R = (1+\aph R)\Delta R + \aph |\nabla R|^2 + \Rc R.
}
To control the behavior of $R(t)$, we will use the maximum principle.
This requires us to analyze solutions of the ODE
$dy/dt = y^2 + \apq y^3$.

\begin{lem}\label{lemA} \mnote{[lemA]}
Suppose $y_0\in \bigl(-\ifap,0\bigr)$. Then the unique solution
$y(t)$ to the initial value problem
\leqn{lemA1}{
\frac{dy}{dt} = y^2 + \apq y^3 \quad : \quad y(0) = y_0
}
exists for all $t\geq 0$ and satisfies
\leqn{lemA2}{
y_0 \leq y(t) < 0 \AND |y(t)| \leq \frac{C_0}{1+t} \quad \forall \; t\geq 0,
}
where $C_0$ is a constant that depends only on $y_0$.
\end{lem}
\begin{proof}
Since $y=0$ and $y=-\ifap$ are the only fixed points of \eqref{lemA1},
$dy/dt>0$ for $-\ifap < y < 0$, and  $y(0)=y_0 \in  \bigl(-\ifap,0\bigr)$,
 it follows that the solution $y(t)$ exists for all $t \geq 0$
and satisfies
\leqn{lemA3}{
y_0 \leq y(t) < 0 \quad \forall \: t \geq 0,
}
and
\leqn{lemA4}{
\lim_{t\rightarrow \infty} y(t) = 0.
}

Next, we observe that \eqref{lemA1} can be integrated to
get
\leqn{lemA5}{
{\apq}
\ln \left(\frac{y_0 (1+{\apq} y(t))}{
y(t)(1+{\apq} y_0)}
\right) + \frac{1}{y_0} - \frac{1}{y(t)} = t.
}
Together, \eqref{lemA3} and \eqref{lemA5} imply that
\leqn{lemA6}{
|y(t)| = \frac{1-{\ifap}|y(t)|\ln(|y(t)|) }{ t + \frac{1}{|y_0|}
 -{\ifap}\ln\left(\frac{|y_0|(1+{\ifap}y(t)) }{1+{\ifap}y_0 }\right) }.
}
But $\lim_{t \rightarrow \infty}
|y(t)|\ln(|y(t)|) = 0$ by \eqref{lemA4}, and so it follows from \eqref{lemA3}
and \eqref{lemA6} that $|y(t)|\leq C_0(1+t)^{-1}$ $(t\geq 0)$ for
some constant $C_0$ that depends only on $y_0$.
\end{proof}

Now, set $C^+_{\Rtl} =  \max_{x\in M}\Rtl(x)$ and
$C^-_{\Rtl} = \min_{x\in M}\Rtl(x)$. Then from equation \eqref{thmp3},
Lemma \ref{lemA}, and the maximum principle (see Theorem 4.4, pg. 96 in
\cite{CK04}), there exists a constant $C_{\Rtl}$ depending
only on $C^{\pm}_{\Rtl}$ such that
\leqn{thmp4}{
C^-_{\Rtl} \leq R(t,x) \AND  |R(t,x)| \leq \frac{C_{\Rtl}}{1+t}
}
for all $(t,x) \in  [0,T)\times M$. Integrating the evolution equation
\eqref{rg2d} in time and applying the inequality \eqref{thmp4} then
yields
\leqn{thmp5}{
|u(t,x)| \lesssim \ln(1+t) \quad \forall \; (t,x) \in  [0,T)\times M.
}
The inequalities \eqref{thmp4}-\eqref{thmp5} together with the formulas
\eqref{cR} show that
\leqn{thmp6}{
(1+t)|\del_t u(t,x)|+|\Deltat u(t,x)| \lesssim 1 \quad \forall \;
(t,x) \in  [0,T)\times M
}

Clearly, the derivative $\del_t u $ satisfies the equation
\eqn{thmp7}{
\del_t(\del_t u) = -D\Rc(u)\cdot \del_t u.
}
By the estimates \eqref{thmp4}-\eqref{thmp6}, this equation is
uniformly parabolic with bounded continuous coefficients. Consequently,
we can apply the Krylov-Safonov estimates \cite{KS} to conclude
the existence of constants $C_T>0$ and $0<\sigma<1$ such that
\leqn{thmp8}{
\norm{\del_t u(t)}_{C^{0,\sigma}(M)} \leq C_T \quad \forall \: t\in [0,T).
}
This implies, increasing $C_T$ if necessary, that
\leqn{thmp9}{
\norm{R(t)}_{C^{0,\sigma}(M)} \leq C_T \quad \forall\: t\in [0,T).
}
Viewing \eqref{rg2u} as an elliptic equation for $u$ with source
term $\del_t u$, the estimates \eqref{thmp4}-\eqref{thmp6}, \eqref{thmp8}
-\eqref{thmp9} allow us to apply Schauder estimates (see Lemma 6.16, pg.
103 of \cite{GT}) to conclude
\eqn{thmp10}{
\norm{u(t)}_{C^{2,\sigma}(M)} \leq C_T \quad \forall\: t\in [0,T).
}
Applying the parabolic continuation principle (see
Proposition 8.1, pg. 338 of \cite{TayIII}), the solution
can be continued for at least a small time past $T$. Thus we
conclude that the solution $u(t)$ exists for all $t\geq 0$.
\end{proof}

\begin{thm} \label{thmm} \mnote{[thmm]}
Suppose $\gt \in \Mcl_-$. Then there exists a smooth one
parameter family of metrics
$g(t)$  for  $-\infty < t \leq 0 $ that satisfy the following:
\begin{itemize}
\item[(i)] $\; g(t)$ solves the $\snd$ order RG equation \eqref{rg}
with $g(0)=\gt$,
\item[(ii)]$\; g(t) \in \Mcl_-$ for all $t \leq 0$, and
\item[(iii)] there exists a constant $C_{\Rtl}$ depending
only on $C^+_{\Rtl} =  \max_{x\in M}\Rtl(x)$ and
$C^-_{\Rtl} = \min_{x\in M}\Rtl(x)$ such that
\eqn{thmm1}{
 R(t,x) \leq C^+_{\Rtl}  \AND  |R(t,x)+\ifap| \leq C_{\Rtl} e^{t}
}
for all $(t,x) \in  (-\infty,0]\times M$.
\end{itemize}
\end{thm}
\begin{proof}
Since the equation \eqref{rg2u} is now backwards parabolic, we instead
consider the forward equation obtained be replacing $t$ with $-t$:
\eqn{thmm2}{
\del_t u = \Rc(u).
}
As in the proof of Theorem \ref{thmp}, we can control the curvature by
analyzing solutions to the ODE $dy/dt = - y^2 -\ifap y^3$.
\begin{lem}\label{lemB} \mnote{[lemB]}
Suppose $y_0\in \bigl(-\ifap,0\bigr)$. Then the unique solution
$y(t)$ to the initial value problem
\leqn{lemB1}{
\frac{dy}{dt} = -y^2 - \apq y^3 \quad : \quad y(0) = y_0
}
exists for all $t\geq 0$ and satisfies
\leqn{lemB2}{
-\ifap < y(t) \leq y_0 \AND |y(t)+\ifap| \leq C_0 e^{-t} \quad \forall \; t\geq 0,
}
where $C_0$ is a constant that depends only on $y_0$.
\end{lem}
\begin{proof}
This proof is essentially the same as the proof of Lemma \ref{lemB} except
for the asymptotics. To see the exponential convergence, we replace $t$ with
$-t$ in the formula \eqref{lemA5} and rearrange to get
\leqn{lemB3}{
|1+{\apq}y(t)| = \frac{|y(t)(1+{\apq}y_0)|}{|y_0|} e^{{\ifap}(1/|y_0|-1/|y(t)|)}
e^{-t}.
}
Since y(t) is bounded by $-\ifap < y(t) \leq y_0 < 0$, the exponential
convergence $|y(t)+\ifap| \leq C_0 e^{-t}$ follows directly from
\eqref{lemB3}.
\end{proof}
Using this lemma and the evolution equation for the scalar curvature
\eqn{lemB4}{
\del_t R = -\Delta R - \Rc R,
}
the proof of Theorem \ref{thmm} follows from a simple adaptation of the
arguments used in the proof of Theorem \ref{thmp}.
\end{proof}

\sect{ccurve}{An eternal solution connecting $\Mcl_-$ to $\Mcl_+$}

Directly from the equation \eqref{rg2d}, it is clear
that the flat metric $\bar{g}$, and the metric $\hat{g}$ with Ricci scalar equal to $-\ifap$ are fixed points for the $\snd$ order RG flow \eqref{rg2d}.
We now show that there exists an eternal
solution that connects the UV fixed point $\hat{g}$ to
the IR fixed point $\bar{g}$. Moreover, we show that this
solution passes through both the sets $\Mcl_\pm$.

To begin, let $\gt$ be a metric with constant negative curvature
\eqn{cc1}{
\Rtl = -1.
}
Next, define
\leqn{cc2}{
g(t,x) := -\frac{1}{y(t)}\gt(x),
}
which implies that the Ricci scalar of $g$ is given by
\leqn{cc3}{
R(t) = y(t).
}
The ansatz \eqref{cc2} is consistent for the $\snd$ order RG flow and leads
to the same equation studied in Lemma \ref{lemA} (i.e. just substitute \eqref{cc3} in \eqref{thmp2}), namely
\leqn{cc4}{
\frac{dy}{dt} = y^2 + \apq y^3.
}
Choosing initial data
\leqn{cc5}{
y(0) = -\itap,
}
it follows immediately from Lemmas \ref{lemA} and \ref{lemB} that the unique solution
$y(t)$ to the initial value problem \eqref{cc4}-\eqref{cc5} exists for all $t\in (-\infty,
\infty)$ and satisfies
\alin{cc6}{
& -\ifap < y(t) < -\itap, \qquad  |y(t)+\ifap|\leq C_- e^{t} \quad \forall \:  t<0,
\intertext{and}
& -\itap < y(t) < 0, \qquad |y(t)| \leq \frac{C_+}{1+t} \quad \forall \: t> 0,
}
for some constants $C_\pm$. In particular, this implies that
\eqn{cc7}{
g(t) \in \Mcl_- \quad \forall \: t <0 \AND g(t) \in \Mcl_+ \quad \forall \: t>0.
}
Moreover, it is clear that
\eqn{cc8}{
\lim_{t\rightarrow -\infty} g(t) = \hat{g} \AND \lim_{t\rightarrow \infty} R(t) = 0.
}
In this sense, the solution $g(t)$ connects the UV fixed point $\hat{g}$ to
the IR fixed point $\bar{g}$.

\sect{disc}{Discussion}

We have shown that the $\snd$ order RG flow admits an eternal solution
that connects a constant negative curvature metric to the flat
metric. Moreover, we have shown that the existence of the eternal solution
is due to the curvature correction term $\aph R_{iklm}R_j{}^{klm}$
to the $\fst$ order flow (i.e. Ricci flow). More specifically,
we demonstrated that the term $\aph R_{iklm}R_j{}^{klm}$
destroys the uniform parabolicity of Ricci flow, and  it is
precisely this lack of uniform parabolicity that allows
for the existence of the eternal solution.

The results of this article show that the lack of uniform
parabolicity of the $\snd$ (and higher) order RG flow
is not necessarily a defect. Instead, the lack of
parabolicity opens up the possibility of constructing
solutions by matching together solutions from neighboring
regions of backwards and forwards parabolicity. We speculate
that this construction could be useful for both physics
and geometry as it
suggests a new method for connecting different geometries (i.e. UV and
IR fixed points) via a geometric flow.

It is important to note that the analysis contained in this paper is
different from that used for 2 dimensional Ricci flow. The main difference is
that control of the scalar curvature $R$, and hence the full curvature tensor,
does not immediately imply, through the use of maximum principles, control on the
higher derivatives of
$R$. This is due to the fully
non-linear nature of the $\snd$ order flow. To gain control on the first derivative on
the scalar curvature, we used the property that $\snd$ order RG flow
reduces to a scalar equation in 2 dimensions which allowed
us to apply the Krylov-Safonov estimates to obtain the desired result.
This prevents the analysis of this paper being extended to higher dimensions.
Consequently, to extend the results of this paper to higher dimensions, a
new method, preferably using standard maximum principles, of obtaining
control on the higher order derivatives of the curvature tensor is required.
Even in 2 dimensions, this would represent a significant improvement over the
analysis contained in this paper.

\bigskip

\noindent \emph{Acknowledgments}

This work was completed while I was visiting the Mittag-Leffler Institute during the
\emph{Geometric, Analysis, and General Relativity program} in the Fall of 2008.
I thank the Institute for its support and hospitality. I also thank J. Isenberg for
helpful suggestions and comments.



\end{document}